\documentclass[10pt,showpacs,preprintnumbers,amsmath,amssymb,amsthm]{revtex4}
\usepackage{units,hyperref,fancyhdr}
\usepackage{bbold,mathpazo}
\usepackage{graphicx}
\usepackage{dcolumn}
\usepackage{bm,bbm,mathtools,esvect}
\usepackage{slashed}
\begin{document}

\title{\sc \Large Quaternionic  Quantum  Particles }
\vspace{2cm}
\author{\sf SERGIO GIARDINO}
\email{sergio.giardino@ufrgs.br}
\affiliation{\vspace{3mm} Departamento de Matem\'atica Pura e Aplicada, Universidade Federal do Rio Grande do Sul (UFRGS)\\
Avenida Bento Gon\c calves 9500, 91509-900 Porto Alegre, RS, Brazil}

\begin{abstract}
\noindent 
Solutions of quaternionic quantum mechanics (QQM) are  difficult to grasp, even in  simple physical situations.
In this article, we provide simple and understandable free particle quaternionic solutions, that can
be easily compared to complex quantum mechanics (CQM). As an application, we study the scattering of quaternionic particles through a scalar step
potential. We also provide a general solution method for the quaternionic Schr\"odinger equation, which can 
be applied to more sophisticated and physically interesting models.
\end{abstract}

\maketitle
\hrule
{\parskip - 0.3mm \footnotesize{\tableofcontents}}
\vspace{1cm}
\hrule
\section{INTRODUCTION\label{I}}%
Quaternionic quantum mechanics (QQM) was conceived by considering complex quantum mechanics (CQM) as a model. However, an important difference 
between these theories concerns the hermiticity of operators: hermitian operators in CQM and anti-hermitian operators in QQM
\cite{Adler:1995qqm}. 
 At first glance, hermiticity seems to be a major difference between the theories, but it is not so. Anti-hermiticity
 cannot be considered a true difference between QQM and CQM because the hermitian Schr\"odinger
equation of CQM turns into an  equation of anti-hermitian operators simply by multiplying of the whole equation 
by the complex unit $i$, and most of the complex hermitian operators are converted to anti-hermitian operators using this procedure.
In terms of phenomenology, both theories calculate expectation values of their operators using identical formulas. The anti-hermitian
formulation of QQM permits the conservation of the probability density current \cite{Adler:1995qqm}, and this is a positive point. 
However, there are drawbacks concerning the anti-hermitian QQM formulation, and the breakdown of Ehrenfest's theorem in QQM
is a serious problem from a physical standpoint \cite{Adler:1985uh,Adler:1988fs,Adler:1995qqm}. 

Another disadvantage of the anti-hermitian QQM is the lack of a set of simple solutions. In CQM, sophisticated and physically interesting models are 
built using a set of simple solutions, which provide the basic understanding of the theory as well.
Of course, there are solutions to QQM, and we can quote some of them by way of example 
\cite{Davies:1989zza,Davies:1992oqq,Ducati:2001qo,Nishi:2002qd,DeLeo:2005bs,Madureira:2006qps,Ducati:2007wp,Davies:1990pm,DeLeo:2013xfa,DeLeo:2015hza,Giardino:2015iia,Sobhani:2016qdp,Procopio:2016qqq,Sobhani:2017yee,DeLeo:2019bcw,Hassanabadi:2017wrt,Hassanabadi:2017jiz},
but they do not have the simplicity that allows them to be compared to either CQM  or classical solutions. 

More recently, an alternative QQM formulation \cite{Giardino:2016nah} has emerged from a quaternionic version of the
Aharonov-Bohm (AB) effect \cite{Giardino:2016xap}, and this formalism has already been applied to scattering states \cite{Mandal:2109qdf}. 
The quaternionic AB-effect has an important difference compared to the usual quantum theories: 
the momentum operator, and consequently the Hamiltonian operator, do not have a defined hermiticity. A consistent theory has been obtained 
by redefining the expectation value and the probability current, so that the probability current is conserved in this 
quaternionic AB-solution. This solution has motivated the development of a quaternionic quantum theory of non-anti-hermitian
operators \cite{Giardino:2016nah}, and this theory has the important advantage of a well-defined classical limit; therefore,
the Ehrenfest theorem is valid in  this framework.  For an arbitrary operator $\mathcal{O}$ and an arbitrary quaternionic
wave function $\Psi$, the expectation value is defined in the non-anti-hermitian theory to be
\begin{equation}\label{I1}
\langle\mathcal O\rangle= \frac{1}{2}\int dx^3\Big[\Psi^\dagger\mathcal{O}\Psi+\big(\Psi^\dagger\mathcal{O}\Psi\big)^\dagger\Big],
\end{equation}
where the dagger denotes the adjoint. This expectation value 
recovers the usual definition when wave functions are complex  and the operators are hermitian, consequently generalizing it. 

In this article, we further develop this non-anti hermitian theory providing a  method for solving the quaternionic Schr\"odinger equation 
and obtaining its simplest example: the free particle. 
This solution has many similarities with the complex wave function, but it also has major differences, mainly the contribution of
the pure quaternionic components of wave function to the probability current.  The quaternionic solutions have constraints that are 
unknown in the complex case, and thus, non trivial solutions are harder to find. On the other hand, quaternionic solutions
present novel features that are unknown in CQM, like a novel time-dependent solution and the possibility 
of controlling the direction of propagation through the relative
amplitude of the complex and pure quaternionic components of the wave function. We have also studied the scattering of quaternionic particles
through a scalar step potential, and the quaternionic particles have higher energies, when compared to the complex case.
 We stress that the solutions presented herein are simple, and it may be immediately verified that they satisfy 
the quaternionic Schr\"odinger equation. Anti-hermitian quaternionic solutions 
\cite{Davies:1989zza,Davies:1992oqq,Ducati:2001qo,Nishi:2002qd,DeLeo:2005bs,Madureira:2006qps,Ducati:2007wp,Davies:1990pm,DeLeo:2013xfa,DeLeo:2015hza,Giardino:2015iia,Sobhani:2016qdp,Procopio:2016qqq}
are much more complicated and subtle, which reduces their interest and potential applicability. The developed general method 
for finding quaternionic solutions is harder when compared to the usual methods for solving the Schr\"odinger equation,
 but the results obtained are illuminating, and we therefore expect it will inspire further research in this area.

This article is organized as follows: in Sections \ref{A1} and \ref{A2} we solve the quaternionic Schr\"odinger equation using a novel method
where a quaternionic solution is obtained from a previously known complex solution. In Section \ref{L} we apply the solution method to find the 
quaternionic free particle, and a few physical aspects of this solution are discussed in Section \ref{P}.
 Section \ref{S} describes the scattering of the quaternionic free particle through a scalar step potential, while 
section \ref{C} rounds off the article with our conclusions and future perspectives.
\section{TIME-DEPENDENT WAVE EQUATIONS\label{A1}}%
In this article, we adopt the wave equation
\begin{equation}\label{a1}
 \hbar\frac{\partial\Psi}{\partial t}\,i=\mathcal{H}\Psi,
\end{equation}
where both the Hamiltonian operator $\mathcal{H}$ and the wave function $\Psi$ are quaternionic. There is a second possibility for (\ref{a1}),
which places the 
complex unit $i$ on the left hand side of (\ref{a1}), so that $\hbar\dot\Psi i\to \hbar i\,\dot\Psi$, where $\dot\Psi$ represents the time derivative
of $\Psi$. We will discuss this possibility in an 
upcoming article, where it can be more conveniently explored. We stress that we use the Schr\"odinger picture of quantum mechanics, where the wave function evolves in time. A Heisenberg picture has been formally obtained in \cite{Giardino:2016nah}, but a more profound understanding 
is a direction for future research.
As in the complex case, we separate the variables of the wave function, so that
\begin{equation}\label{a2}
 \Psi(\bm x,\,t)=\Phi(\bm x)\Lambda(t).
\end{equation}
$\Phi$ and $\Lambda$ are quaternionic functions, and we additionally impose $\Lambda$  as an unitary quaternion with
\begin{equation}\label{a3}
 \Lambda=\cos\Xi\, e^{iX}+\sin\Xi\, e^{i \Upsilon}j,\qquad\mbox{so that}\qquad \Lambda\Lambda^*=1,
\end{equation}
where $\Lambda^*$ is the quaternionic conjugate. 
Of course, $\Xi,\,X$ and $\Upsilon $ are time-dependent real functions and $j$ is the anti-commuting quaternionic complex unit, which obeys
$ij=-ji$ and $j\,^2=-1$. We adopt the symplectic notation for quaternionic numbers ($\mathbb{H}$), where $q\in\mathbb{H}$ is written
\begin{equation}\label{a4}
 q=z +\zeta j\qquad\mbox{with}\qquad z,\,\zeta\in\mathbb{C}.
\end{equation}
$\Lambda$ is already written in symplectic notation (\ref{a3}), and $\Phi$ will be expressed in the same terms. An unitary quaternion is the natural generalization of the complex
exponential function that is the time-dependent part of a complex wave function, and both functions have
oscillatory properties that are desirable in quantum theories. Real exponentials are not deployed in CQM, nor will they be considered
here, either. In order to separate the time variable in the wave function, we use (\ref{a2}) and (\ref{a3}) in the wave equation
(\ref{a1}), and we further impose
\begin{equation}\label{a5}
 \dot\Lambda i\Lambda^*=\kappa.
\end{equation}
The dot denotes a time derivative, and $\kappa=\kappa_0+\kappa_1j$ is a quaternionic 
separation constant with $\kappa_0$ and $\kappa_1$ complex. Thus, from (\ref{a5}) we get
\begin{equation}\label{a6}
-i\,\dot\Xi\,\sin 2\,\Xi-\dot X\,\cos^2\Xi+\dot\Upsilon\,\sin^2\,\Xi-\frac{i}{2}
\left[\sin 2\,\Xi\, e^{i(X+\Upsilon)}\right]^{\bm\cdot} j=\kappa\,.
\end{equation}
We separate it into two parts according to the dependence on the complex unit $j$, and the pure quaternionic part gives
\begin{equation}\label{a7}
-\frac{i}{2}\sin 2\,\Xi\, e^{i(X+\Upsilon)}=\kappa_1 t+C.
\end{equation}
where $C$ is an integration constant. The sine is finite, and consequently $\kappa_1=0$. The simplest solution is thus
\begin{equation}\label{a8}
 \Lambda(t)=\Lambda_0\exp\left[-i\frac{\mathcal{E}}{\hbar}t\right],
\end{equation}
where $\mathcal{E}$ is the energy and $\Lambda_0$ is an unitary quaternionic constant. (\ref{a8}) is simply the 
CQM time-dependent solution with a constant quaternionic phase. A more general solution is obtained for
\begin{equation}\label{a9}
\dot\Xi=0\qquad\mbox{and}\qquad\dot X=-\dot\Upsilon=-\frac{\mathcal{E}}{\hbar},
\end{equation}
so that
\begin{equation}\label{a10}
 \Lambda(t)=\Lambda_0\left\{\cos\Xi \,\exp\left[-i\frac{\mathcal{E}}{\hbar}t\right]+ 
\sin\Xi\, \exp\left[i\left(\frac{\mathcal{E}}{\hbar}t+\tau_0\right)\right]j\right\},
\qquad\qquad \dot\Lambda i\Lambda^*=\frac{\mathcal{E}}{\hbar},
\end{equation}
and $\tau_0$ is a real constant that comes from the difference between the integration constants of $X$ and $Y$ in (\ref{a9}). Equation (\ref{a10}) presents an amazingly simple quaternionic result, which recovers the complex time-dependent quantum 
solution when $\Lambda_0$ is complex and $\Xi=0$, and thus generalizes the complex case. In the next section, we proceed accordingly to find 
 time-independent quantum quaternionic solutions that are simple and new like (\ref{a10}).

\section{TIME-INDEPENDENT WAVE EQUATIONS\label{A2}}

From the previous section, the time-independent Schr\"odinger equation is
\begin{equation}\label{A8}
 \mathcal{H}\Phi=\mathcal{E}\Phi,
\end{equation}
and we notice that the hamiltonian operator $\mathcal{H}$ is not supposed to be either hermitian or anti-hermitian, only the energy 
$\mathcal{E}$ is supposed to be real. In general, the Hamiltonian operator is
\begin{equation}
\mathcal{H}=-\frac{\hbar^2}{2m}\nabla^2+V.
\end{equation}
In general, the scalar potential $V$ can be quaternionic, but in this article we assume it to be real. We propose the spatial wave function
\begin{equation}\label{A9}
\Phi=\phi\,\lambda,
\end{equation}
where $\lambda$ is a quaternionic function and $\phi$ is a time-independent complex solution of 
Schr\"odinger equation with energy $E$, so that $\mathcal{H}\phi=E\phi$. Using (\ref{A8}-\ref{A9}) and the identity
$\;\nabla^2(\phi\lambda)=(\nabla^2\phi)\lambda +2\bm\nabla\phi\bm{\cdot\nabla}\lambda+\phi\nabla^2\lambda,\;$ we get
\begin{equation}\label{A10}
\phi \,\nabla^2\lambda+2\bm\nabla\phi\bm{\cdot\nabla}\lambda=\frac{2m}{\hbar^2}\big(E-\mathcal{E}\big)\phi\lambda.
\end{equation}
It is important to notice that (\ref{A10}) is valid for every potential $V$, what demonstrates the generality of the method. We can make further progress decomposing the quaternionic function, such that
\begin{equation}\label{A1100}
 \lambda=\rho K,\qquad\mbox{where}\qquad |\lambda|=\rho,\qquad K=\cos\Theta\, e^{i\Gamma}+\sin\Theta\, e^{i \Omega}j\,,
\qquad\mbox{and}\qquad KK^*=1;
\end{equation}
where $\rho,\,\Theta,\,\Gamma$ and $\Omega$ are real functions. We then turn (\ref{A10}) into
\begin{equation}\label{A11}
\frac{1}{\rho}\left[ \big(\bm\nabla+\frac{2}{\phi}\bm\nabla\phi\big)\bm{\cdot\nabla}\rho\right] K+
\frac{2}{\rho\phi}\bm\nabla(\rho\phi)\bm{\cdot\nabla}K+\nabla^2 K=\frac{2m}{\hbar^2}\big(E-\mathcal{E}\big)K.
\end{equation}
Defining
\begin{equation}\label{A110}
 \bm\nabla K=\bm p\, e^{i\Gamma}+\bm q\, e^{i\Omega}j\qquad\mbox{and}\qquad \nabla^2 K=u\, e^{i\Gamma}+ v\, e^{i\Omega}j,
\end{equation}
where
\begin{align}
& \bm p=-\sin\Theta\,\bm\nabla\Theta+i\cos\Theta\,\bm\nabla\Gamma,\qquad
\bm q= \cos\Theta\,\bm\nabla\Theta+i\,\sin\Theta\,\bm\nabla\Omega,\nonumber\\
& u= -\cos\Theta\,\Big(\,\big|\bm\nabla\Gamma\big|^2+\big|\bm\nabla\Theta\big|^2\,\Big)-\sin\Theta\,\nabla^2\Theta
+i\Big(\cos\Theta\,\nabla^2\Gamma-2\sin\Theta\,\bm\nabla\Gamma\bm{\cdot\nabla}\Theta\Big)\label{A12}\\
 & v=-\sin\Theta\,\Big(\,\big|\bm\nabla\Omega\big|^2+\big|\bm\nabla\Theta\big|^2\,\Big)+\cos\Theta\,\nabla^2\Theta
+i\Big(\sin\Theta\,\nabla^2\Omega+2\cos\Theta\,\bm\nabla\Omega\bm{\cdot\nabla}\Theta\Big),\nonumber
\end{align}
we can split (\ref{A11}) into its complex and quaternionic parts, respectively
\begin{align}
& \label{A13}
\frac{1}{\rho}\big(\bm\nabla+\frac{2}{\phi}\bm\nabla\phi\big)\bm{\cdot\nabla}\rho+
\frac{2}{\rho\phi}\bm\nabla(\rho\phi)\bm\cdot\frac{\bm p}{\cos\Theta}+\frac{u}{\cos\Theta}=\frac{2m}{\hbar^2}\big(E-\mathcal{E}\big) \\
& \label{A14}
\frac{1}{\rho}\big(\bm\nabla+\frac{2}{\phi}\bm\nabla\phi\big)\bm{\cdot\nabla}\rho+
\frac{2}{\rho\phi}\bm\nabla(\rho\phi)\bm\cdot\frac{\bm q}{\sin\Theta}+\frac{v}{\sin\Theta}=\frac{2m}{\hbar^2}\big(E-\mathcal{E}\big)
\end{align}
After defining, 
\begin{equation}\label{A15}
 \frac{1}{\rho}\left(\bm\nabla+\frac{2}{\phi}\bm\nabla\phi\right)\bm{\cdot\nabla}\rho=\mathcal{Z}_0,\qquad
\frac{2}{\rho\phi}\bm\nabla(\rho\phi)\bm\cdot\frac{\bm p}{\cos\Theta}=\mathcal{Z}_1\qquad\mbox{and}\qquad
\frac{2}{\rho\phi}\bm\nabla(\rho\phi)\bm\cdot\frac{\bm q}{\sin\Theta}=\mathcal{Z}_2,
\end{equation}
where $\mathcal{Z}_0,\,\mathcal{Z}_1$ and $\mathcal{Z}_2$ are complex functions,
we separate (\ref{A13}-\ref{A14}) into real components, so that
\begin{align}
\label{A16}
& \Re(\mathcal{Z}_0+\mathcal{Z}_1)-|\bm\nabla\Gamma\big|^2-\big|\bm\nabla\Theta\big|^2-\tan\Theta\nabla^2\Theta
=\,\frac{2m}{\hbar^2}\big(E-\mathcal{E}\big) \\
\label{A17}
& \Re(\mathcal{Z}_0+\mathcal{Z}_2)-|\bm\nabla\Omega\big|^2-\big|\bm\nabla\Theta\big|^2+
\cot\Theta\nabla^2\Theta=\,\frac{2m}{\hbar^2}\big(E-\mathcal{E}\big) \\
\label{A18}
&\Im(\mathcal{Z}_0+\mathcal{Z}_1)+\Big(\bm\nabla-2\tan\Theta\bm\nabla\Theta\Big)\bm{\cdot\nabla}\Gamma=\,0\\
\label{A19}
&\Im(\mathcal{Z}_0+\mathcal{Z}_2)+\Big(\bm\nabla+2\cot\Theta\bm\nabla\Theta\Big)\bm{\cdot\nabla}\Omega=\,0
\end{align}
where $\Re(\mathcal{Z})$ and $\Im(\mathcal{Z})$ are respectively the real and the imaginary components of a complex $\mathcal{Z}$. 
Thus, every time-independent complex wave function $\phi$ that satisfies the Schr\"odinger equation with real energy
$E$ has a correspondent quaternionic time-independent wave function. When multiplied by a quaternionic time-independent function 
$\lambda=\rho K$, the complex wave function generates
the quaternionic wave function $\Phi=\phi\lambda$, which satisfies  (\ref{A15}-\ref{A19}) and the quaternionic wave equation (\ref{A8}).
To the best of our knowledge, this is the simplest and most general method of finding non-relativistic quaternionic wave
functions. In the following sections we will look for the simplest quaternionic functions that satisfy these conditions and ascertain
several physical properties.
\section{THE QUATERNIONIC PARTICLE SOLUTIONS\label{L}}
The wave function of a complex free quantum particle is represented by a complex exponential, and thus the amplitude of the wave function is constant. 
In analogy to the complex case, we suppose that the quaternionic radius $\rho$ is constant, and therefore
\begin{equation}\label{L1}
 \bm\nabla\rho=0,
\end{equation}
In this case, the quaternionic wave function is generated from $\phi$ through a quaternionic geometric phase. A complete
investigation of such a quaternionic Berry phase is interesting enough to be the subject of a separate article, which we will do at a latter date. 
For our purposes, we further simplify our problem if
\begin{equation}\label{L2}
 \bm\nabla\phi\bm{\cdot\nabla}\Gamma=0,\qquad\bm\nabla\phi\bm{\cdot\nabla}\Omega=0,\qquad\bm\nabla\phi\bm{\cdot\nabla}\Theta=0,
\end{equation}
and then two possible simple solutions emerge. Let us see the first.
\subsection{$\nabla^2\Theta= 0$}
Using this choice, 
\begin{equation}\label{L5}
\Gamma=\bm{\gamma\cdot x}+\Gamma^{(0)},\qquad \Omega=\bm{\omega\cdot x}+\Omega^{(0)},\qquad \Theta=\bm{\theta\cdot x}+\Theta^{(0)},
\end{equation}
with $\bm{\gamma,\,\omega}$ and $\bm\theta$ constant real vectors and $\Gamma^{(0)},\,\Omega^{(0)}$ and $\Theta^{(0)}$ real constants. We 
thus obtain several constraints involving the vector norms,
\begin{equation}\label{L6}
|\bm\gamma|^2=|\bm\omega|^2,\qquad
|\bm\gamma|^2+|\bm\theta|^2=\frac{2m}{\hbar^2}(\mathcal{E}-E),
\end{equation}
and orthogonality constraints as well, 
\begin{equation}\label{L7}
\bm{\theta\cdot\gamma}=0,\qquad\bm{\theta\cdot\omega}=0,\qquad\bm\nabla\phi\bm{\cdot\theta}=0,\qquad\bm\nabla\phi\bm{\cdot\gamma}=0,
\qquad\mbox{and}\qquad\bm\nabla\phi\bm{\cdot\omega}=0.
\end{equation}
In the same fashion that $\phi$ is a linear combination of two complex exponentials according to the sign of the exponent, 
the quaternionic solution has four possibilities, and hence the most general wave function is
\begin{align}\nonumber
& \Phi=
\phi(\bm x)\Big[\left(\cos\Theta\, e^{\,i\Gamma}+\sin\Theta\, e^{\,i\Omega}j\right)Q_1+
\left(\cos\Theta\,e^{\,i\Gamma}+\sin\Theta\,e^{-i\Omega}j\right)Q_2+\\
&\label{L8}
+\left(\cos\Theta\, e^{-i\Gamma}+\sin\Theta\, e^{\,i\Omega}j\right)Q_3+
\left(\cos\Theta\,e^{-i\Gamma}+\sin\Theta\,e^{-i\Omega}j\right)Q_4
\Big],
\end{align}
where $Q_1,\,Q_2,\,Q_3$ and $Q_4$ are arbitrary quaternionic constants. We stress that $\,\pm\Gamma\,$ and $\,\pm\Omega\,$ satisfy (\ref{L2}), and
thus (\ref{L8}) entertains all the possibilities for the wave function. We can make
$\phi$ constant, so that $E=0$, and thus obtain the simplest solution of the case, a truly quaternionic free particle. However, we 
remember that every complex wave function may generate this kind of quaternionic solution. Furthermore, we stress that (\ref{L8}) is very
different from the complex case. The function $\Theta$ enables the wave function with a novel oscillation, totally unknown in CQM.
\subsection{$\nabla^2\Theta\neq 0$}
Let us choose
\begin{equation}\label{L9}
 \Gamma=\bm{\gamma\cdot x}+\Gamma^{(0)},\qquad\Omega=\bm{\omega\cdot x}+\Omega^{(0)},
\end{equation}
and the vector constraints
\begin{equation}\label{L10}
 \bm\nabla\phi\bm{\cdot\gamma}=0,\qquad\bm\nabla\phi\bm{\cdot\omega}=0,\qquad\bm\nabla\phi\bm{\cdot\nabla}\Theta=0,\qquad\bm\nabla\Theta\bm{\cdot\gamma}=0
\qquad\mbox{and}\qquad\bm\nabla\Theta\bm{\cdot\omega}=0.
\end{equation}
The equations to be solved are
\begin{equation}\label{L11}
|\bm\nabla\Theta|^2+|\bm\omega|^2\sin^2\Theta+|\bm\gamma|^2\cos^2\Theta=\frac{2m}{\hbar^2}(\mathcal{E}-E)\qquad\mbox{and}\qquad 
 \nabla^2\Theta+\Big(|\bm\gamma|^2-|\bm\omega|^2\Big)\sin\Theta\cos\Theta=0.
\end{equation}
If we apply the gradient operator to the $|\bm\nabla\Theta|^2$ equation, we obtain the $\nabla^2\Theta$ equation with a changed sign,
and consequently $\nabla^2\Theta=0$, recovering the previous case. Thus there is no non-trivial solution to this situation.

\section{THE FREE QUATERNIONIC PARTICLE\label{P}}

The complex free particle solution enables important quantum models to be developed, such as the infinite well, the finite
well, scattering phenomena and wave packets. In the previous section, we saw that this complex wave function has a quaternionic
generalization. In this section, we will discuss this mathematical solution in physical terms.
Accordingly, we will make use of the probability current defined in \cite{Giardino:2016nah}, namely
\begin{equation}\label{P1}
 \bm J=\frac{1}{2m}\Big[\Phi^*\bm{\hat p}\Phi+\big(\Phi^*\bm{\hat p}\Phi\big)^*\Big],
\end{equation}
and the quaternionic momentum $\bm{\hat p}$ is defined in \cite{Giardino:2016nah} as
\begin{equation}\label{P2}
 \bm{\hat p}=-\hbar(\bm\nabla|i), \qquad\mbox{so that}\qquad \bm{\hat p}\Phi=-\hbar\bm\nabla\Phi i.
\end{equation}
Using (\ref{A9}) and (\ref{A110}), we get the probability current of the quaternionic wave function to be
\begin{equation}\label{P200}
\bm J=\rho^2\cos 2\Theta\,\bm J_0+\frac{\hbar}{m}\rho^2|\phi|^2\left(\cos^2\Theta\,\bm\nabla\Gamma-\sin^2\Theta\,\bm\nabla\Omega\right),
\end{equation}
where $\bm J_0$ is the CQM probability current. 
By way of example, let us calculate the probability current of a quaternionic wave function (\ref{L8}) where $Q_1=1$
and $Q_2=Q_3=Q_4=0$, and the complex wave function is the free particle
\begin{equation}\label{P0}
\phi(\bm x)=A_1\,e^{\,i\bm{k\cdot x}}+A_2\,e^{-i\bm{k\cdot x}},\qquad\mbox{where}\qquad |\bm k|^2=\frac{2mE}{\hbar^2},
\end{equation}
where $A_1$ and $A_2$ are complex constants.  From (\ref{L8}) and (\ref{P200}), we get
\begin{equation}\label{P4}
 \bm J=\frac{\hbar}{m}\left[\cos 2\Theta\left(|A_1|^2-|A_2|^2\right)\bm k+
|\phi|^2\left(\cos^2\Theta\,\bm\nabla\Gamma-\sin^2\Theta\,\bm\nabla\Omega\right)\right].
\end{equation}
$\bm J$ satisfies the continuity equation
\begin{equation}\label{P5}
\frac{\partial \varrho}{\partial t}+ \bm\nabla\cdot \bm J=0,
\end{equation}
where $\varrho=|\Psi|^2$ is the probability density and thus there is no gain and no loss of probability in the free particle solution. As expected, the complex result
is recovered if $\bm\nabla\Gamma=\bm 0$ and $\Theta=0$. There is no probability
flux along the $\bm\nabla\Theta$ direction, and consequently no momentum is carried along this direction, however, the particle is allowed to
move along the $\bm\nabla\Theta$ direction. Along the directions orthogonal to $\bm\nabla\Theta$, the
$\Theta$ function is a simple parameter of the solution, and it may be made constant without affecting the
probability flux and, of course, on the momentum expectation value.

Another feature involves the contributions for the probability flux due to $\bm\nabla\Gamma$ and $\bm\nabla\Omega$. If 
these vectors are collinear and have a common positive direction, 
 $\bm\nabla\Gamma$ increases $\bm J$ and $\bm\nabla\Omega$ attenuates the flux. 
This fact may be interpreted in terms of anti-particles that move in the opposite direction of the particle, and this interpretation has been 
made for relativistic quaternionic solutions \cite{DeLeo:2013xfa,DeLeo:2015hza,Giardino:2015iia}. However, this may not necessarily be the case. 
Another interpretation is that $(\cos^2\Theta\,\bm\nabla\Gamma-\sin^2\Theta\,\bm\nabla\Omega)$ contributes the probability flux of a single particle.
The current attenuates and even inverts depending on the relation between the amplitudes of the complex and quaternionic parts of the wave 
function. This interpretation does not require the existence of anti-particles, and it is in agreement with the complex case. 
Two simple  examples illustrate the situation.
\begin{equation}\label{P6}
\Phi_1=\cos\Theta\, e^{i\bm{k\cdot x}}+\sin\Theta\, e^{-i\bm{k\cdot x}}\,j\qquad\mbox{and}\qquad
\Phi_2=e^{i\bm{k\cdot x}}\big(\cos\Theta +\sin\Theta\, j\,\big), 
\end{equation}
where $\Theta$ is a real constant and $k$ is a real vector. The probability currents are
\begin{equation}\label{P7}
 \bm J_1=\frac{\hbar}{m}\bm k\qquad \mbox{and} \qquad \bm J_2= \frac{\hbar}{2m}\cos2\Theta\,\bm k. 
\end{equation}
The intensity of $\bm J_1$ is identical to the intensity of the probability current generated by the complex particle (\ref{P0}) when $A_2=0$, and
 we interpret $\Phi_1$ as a single particle wave function. On the other hand, $\Phi_2$ generates a probability flux whose intensity depends
on $\Theta$, and even $\bm J_2=\bm 0$ if $\Theta =\pi/4$. Accordingly, we can understand $\bm J_2$ as a single particle probability current,
and $\Theta$ determines the intensity of the probability flux. Although there is neither
probability current nor momentum along the $\bm\nabla\Theta$ direction, the value of the angle affects the direction and the intensity 
of the flux, and thus $\Theta$ is in fact a relevant physical variable whose precise meaning must be determined on a case-by-case basis.

\section{\label{S}THE STEP POTENTIAL}
The quaternionic solutions we examined in the previous section may be written as
\begin{equation}\label{S1}
 \Phi=\phi K,
\end{equation}
where $\phi$ is a complex wave function and $K$ is a unit quaternion, and can be understood as a complex solution with
a quaternionic phase. This kind of solution can hardly generate physically novel solutions because geometric phases generate observable
effects only in interaction processes. Conversely, we observed that the single particles (\ref{P6}) present differences compared to the 
complex cases, even without inter-particle interactions. Let us go further to observe whether quaternionic effects may happen to single particle
systems, like the scattering of quaternionic free particles by the scalar step potential
\begin{equation}\label{S2}
 V=\left\{
\begin{array}{ccc}
0 & \qquad\; \mbox{for}\qquad  x<0,&\;\;\;\;\;\mbox{region I} \\
V_0 & \qquad \mbox{for}\qquad  x\ge 0, &\qquad\mbox{region II},
\end{array}
\right.
\end{equation}
where $V_0$ is a real positive constant and the potential $V$ splits the three-dimensional space into
two parts bordered by the $Oyz$ plane. We propose the wave function
\begin{align}\nonumber
\Phi_I&=\cos\Theta_k e^{i(\bm{k\cdot x}+\bm\gamma_k^\perp\bm{\cdot x})}+\sin\Theta_k e^{i(-\bm{k\cdot x}+\bm\omega_k^\perp\bm{\cdot x})}\,j
+R\left[\cos\Theta_q e^{i(-\bm{q\cdot x}+\bm\gamma_q^\perp\bm{\cdot x})}+\sin\Theta_q e^{i(\bm{q\cdot x}+\bm\omega_q^\perp\bm{\cdot x})}\,j\right]
 \\ \label{S3}
\Phi_{II}&=T\left[\cos\Theta_p e^{i(\bm{p\cdot x}+\bm\gamma_p^\perp\bm{\cdot x})}+\sin\Theta_p e^{i(-\bm{p\cdot x}+\bm\omega_p^\perp\bm{\cdot x})}\,j\right]
\end{align}
with $R$ and $T$ complex constants and $\bm k,\,\bm q,\,\bm p,\,\bm\gamma^\perp_a$ and $\bm\omega^\perp_a$ real vectors for $a=k,\,q,\,p$. 
We also suppose that $\Theta_a$ are constants and adopt that an arbitrary vector $\bm v$ splits into components according to 
\begin{equation}\label{S4}
 \bm v=\bm v^\parallel +\bm v^\perp,
\end{equation}
where $\bm v^\parallel$ is the component of $\bm v$ that is parallel to $\bm k$ and $\bm v^\perp$ is the component of $\bm v$ that is 
normal to $\bm k$. Schr\"odinger's equation permit us to get
\begin{equation}\label{S5}
|\bm k|^2+|\bm\gamma_k^\perp|^2=|\bm q|^2+|\bm\gamma_q^\perp|^2=\frac{2m}{\hbar^2}\mathcal{E}\qquad\mbox{and}\qquad |\bm p|^2+|\bm\gamma_p^\perp|^2=
\frac{2m}{\hbar^2}\left(\mathcal{E}-V_0\right),
\end{equation}
where it has been used that
\begin{equation}\label{S6}
|\bm\gamma_a^\perp|^2=|\bm\omega_a^\perp|^2\qquad\mbox{for}\qquad a=k,\,p,\,q.
\end{equation}
There is a set of boundary conditions at the point of incidence $\bm x_0=(0,\,y_0,\,z_0)$, but we can set $x_0=(0,\,0,\,0)$ without loss of
generality. Considering the continuity of the wave function, 
\begin{align}\label{S7}
\Phi_I (\bm x_0)=\Phi_{II}(\bm x_0)\qquad&\Rightarrow\qquad 
\left\{
\begin{array}{ll}
&\cos\Theta_k^{(0)}+R\cos\Theta_q^{(0)}=T\cos\Theta_p^{(0)}\\
&\sin\Theta_k^{(0)}+R\sin\Theta_q^{(0)}=T\sin\Theta_p^{(0)}
\end{array}
\right.
\\ \label{S8}
\bm\nabla\Phi_I^\parallel (\bm x_0)=\bm\nabla\Phi_{II}^\parallel(\bm x_0)\qquad &\Rightarrow\qquad 
\left\{
\begin{array}{ll}
&\bm k\cos\Theta_k^{(0)}-R\bm q\cos\Theta_q^{(0)}=T\bm p\cos\Theta_p^{(0)}\\
&-\bm k\sin\Theta_k^{(0)}+R\bm q\sin\Theta_q^{(0)}=-T\bm p\sin\Theta_p^{(0)}.
\end{array}
\right.
\end{align}
Considering $\bm k,\,\bm q$ and $\bm p$ collinear vectors, we obtain
\begin{equation}\label{S9}
 |T|^2=\frac{|\bm k+\bm q|^2}{|\bm p+\bm q|^2}\qquad\mbox{and}\qquad |R|^2=\frac{|\bm k-\bm p|^2}{|\bm p+\bm q|^2}.
\end{equation}
We obtain the simplest solution imposing real coefficients $R$ and
\begin{equation}\label{S12}
|\bm k|=|\bm q|.
\end{equation}
Furthermore, (\ref{S5}) requires
\begin{equation}
 |\bm\gamma_k^\perp|^2=|\bm\gamma_q^\perp|^2.
\end{equation}
We notice that $\bm\gamma^\perp_a=\bm 0$ recovers the probability density of the complex case, but the wave function is still quaternionic. A
complex limit is only recovered if $\Theta_a=0$, and we 
conclude that the wave function (\ref{S3}) generalizes the complex case because the physical results of the complex case are recovered within
a limit, even though the wave function is still quaternionic. We can hypothesize that complex solutions may have a 
quaternionic counterpart with identical physical predictions.  Yet, the analysis of the quaternionic step 
demands the boundary conditions along the $\bm k^\perp$ direction
\begin{equation}\label{S121}
\bm\nabla\Phi_I^\perp (\bm 0)=\bm\nabla\Phi_{II}^\perp(\bm 0)\qquad \Rightarrow\qquad 
\left\{
\begin{array}{cc}
&\bm\gamma_k^\perp\cos\Theta_k^{(0)}+R\bm\gamma_q^\perp\cos\Theta_q^{(0)}=T\bm\gamma_p^\perp\cos\Theta_p^{(0)}\\
&\bm\omega_k^\perp\sin\Theta_k^{(0)}+R\bm\omega_q^\perp\sin\Theta_q^{(0)}=T\bm\omega_p^\perp\sin\Theta_p^{(0)}.
\end{array}
\right.
\end{equation}
To finally solve the problem, let us suppose that the $\bm\gamma_a$ vectors are collinear, and so are the $\bm\omega_a$ vectors. We stress that
$\bm\gamma^\perp_a$ and $\bm\omega^\perp_a$ are not necessarily collinear, although they have identical magnitudes. We also remember that 
$R$ and $T$ satisfy the CQM relation 
\begin{equation}\label{S120}
|\bm k|-|\bm q|R=|\bm p|T. 
\end{equation}
Comparing (\ref{S121}) with (\ref{S120}), we obtain
\begin{equation}\label{S13}
\frac{|\bm\gamma_q|\cos\Theta_q^{(0)}}{|\bm\gamma_k|\cos\Theta_k^{(0)}}=
\frac{|\bm\omega_q|\sin\Theta_q^{(0)}}{|\bm\omega_k|\sin\Theta_k^{(0)}}=-1
\qquad\mbox{and}\qquad
\frac{|\bm\gamma_p|\cos\Theta_p^{(0)}}{|\bm\gamma_k|\cos\Theta_k^{(0)}}=
\frac{|\bm\omega_p|\sin\Theta_p^{(0)}}{|\bm\omega_k|\sin\Theta_k^{(0)}}=\frac{|\bm p|}{|\bm k|}.
\end{equation}
(\ref{S13}) permits us say that
\begin{equation}\label{S14}
 \sin^2\Theta_k^{(0)}=\sin^2\Theta_q^{(0)}=\sin^2\Theta_p^{(0)}.
\end{equation}
and thus,
\begin{equation}\label{S15}
 \frac{|\bm p|^2}{|\bm k|^2}=\frac{|\bm\gamma^\perp_p|^2}{|\bm\gamma^\perp_k|^2}=1-\frac{V_0}{\mathcal{E}}.
\end{equation}
The (\ref{S15}) relation, between the incoming and transmitted momenta, is equivalently obeyed in the complex case, and thus
the quaternionic particle (\ref{S3}) is possibly the simplest quaternionic generalization of the complex scattering. 
The solution depends only on the parameters of the incident particle: the energy $\mathcal{E}$, the transverse momentum $\bm\gamma_k$ and on the 
intensity of the scalar potential $V_0$ and $\Theta_k$ angle. 
We observe that all the components of probability $\bm J$ are collinear, but the flux is not necessarily normal to the $Oxy-$plane. The direction
of the incident probability flow is determined by $\Theta_k,\,\bm\gamma_k$ and $\bm\omega_k$, while the only effect of $V_0$ is the intensity
of the transmitted momentum $\bm p$. The potential does not change the direction of the transmitted and reflected particles. This means
that the system does not behave like a light beam, and consequently does not satisfy some version of the Snell law. This remark is important
because this a behavior has been observed in anti-hermitian QQM \cite{Ducati:2013:qsl}, and the reasons for this difference must be 
investigated.

\section{CONCLUSION\label{C}}

In this article we presented a general method for finding quaternionic quantum solutions. The quaternionic time-dependet wave function
generalizes the CQM result with a unitary quaternion, whereas  starting points of the time-independent solution	
are  CQM-wave functions $\phi$ that generate quaternionic solutions when multiplied to a quaternion function $\lambda$, 
and hence the quaternionic solutions are such that $\Phi=\phi\lambda$. The solution method does not suppose hermiticity for operators, and thus
the results are out of the main stream of QQM, which is mainly anti-hermitian. Our approach, however, has the advantage of a well-behaved classical
limit, something not accomplished within the anti-hermitian framework. 

The method has been applied to obtain the simplest quantu solution: the free particle. The quaternionic free particle
has a limit whose physical properties are identical to the complex quantum free particle. However, the wave function has novel oscillations,
and this is evidence that our approach must be considered as an alternative quaternionic generalization of quantum mechanics. 

The scattering of a quaternionic particle through a scalar step potential has also presented many similarities with the complex case. The 
parameters of the quaternionic case may regulate the intensity and direction of the probability flow, something unobserved in the 
complex case. The solution is very simple compared to the anti-hermitian quaternionic cases,  for example \cite{Madureira:2006qps}, and then
we hope that this simplicity will inspire further research in the field.

We can outline several perspectives for future research. Many mathematical questions that are currently studied in anti-hermitian QQM 
\cite{Sabadini:2017wha} can be addressed to the non-anti-hermitian formalism used in the present article.
On another side, throughout the text we have observed that there is a second possibility to the wave
equation (\ref{a2}), where $i$ is placed on the left hand side of $\dot\Psi$, and also the study of quaternionic geometric phases, which
have already been started with the quaternionic AB-effect \cite{Giardino:2016xap}, but that may deserve a more specific treatment. In principle,
every result of quantum mechanics is at risk of being revised using the non-anti-hermitian formalism, and therefore, the field of
research is vast. We hope that it can be thoroughly investigated in the future.

\section*{ACKNOWLEDGEMENTS}
Sergio Giardino is grateful for the hospitality at the Institute of Science and Technology of Unifesp in S\~ao Jos\'e dos Campos.

%
%
%
%

\bibliographystyle{unsrt} 
\bibliography{bib_qfree}

\end{document}